%% file: Tantalum.tex
\documentclass[twocolumn,superscriptaddress,nofootinbib]{revtex4-1}

\usepackage{comment}
\usepackage{slashed}
\usepackage{latexsym}
\usepackage{graphics}
\usepackage{bm}
\usepackage{color}
\usepackage{amsmath}
\usepackage{amssymb}
\usepackage{feynmp-auto}
\usepackage{graphicx}
\usepackage{subcaption}
\usepackage{slashed}
\usepackage{lineno}

\usepackage{color,hyperref}
\hypersetup{colorlinks,breaklinks,
 linkcolor=blue,urlcolor=blue,
 anchorcolor=blue,citecolor=blue}

\usepackage{tikz}
\usetikzlibrary{shapes.geometric,decorations.fractals,shadows,shapes.multipart,arrows,snakes,positioning,arrows}
\usetikzlibrary{decorations.pathmorphing,decorations.markings}
\usetikzlibrary{calc,intersections}
\begin{document}
  \input{commands}
\title{Search for dark matter induced de-excitation of \texorpdfstring{$^{180}$Ta$\rm ^m$}{Lg}}
\author{Bj\"{o}rn Lehnert}
\email{bjoernlehnert@lbl.gov}
\affiliation{Nuclear Science Division, Lawrence Berkeley National Laboratory, Berkeley, CA 94720}

\author{Harikrishnan Ramani}
\email{hramani@berkeley.edu}
\affiliation{Berkeley Center for Theoretical Physics, Department of Physics,
University of California, Berkeley, CA 94720, USA}
\affiliation{Theoretical Physics Group, Lawrence Berkeley National Laboratory, Berkeley, CA 94720}

\author{Mikael Hult}
\affiliation{European Commission, JRC-Geel, Retieseweg 111, B-2440 Geel, Belgium}

\author{Guillaume Lutter}
\affiliation{European Commission, JRC-Geel, Retieseweg 111, B-2440 Geel, Belgium}

\author{Maxim Pospelov}
\affiliation{Perimeter Institute for Theoretical Physics, Waterloo, ON N2J 2W9, Canada}
\affiliation{Department of Physics and Astronomy,
University of Victoria, Victoria, BC V8P 5C2, Canada}
\author{Surjeet Rajendran}
\affiliation{Berkeley Center for Theoretical Physics, Department of Physics,
University of California, Berkeley, CA 94720, USA}
\author{Kai Zuber}
\affiliation{Institut f\"{u}r Kern- und Teilchenphysik, Technische Universit\"{a}t Dresden, Germany}

\begin{abstract}

Weak-scale dark matter particles, in collisions with nuclei, can mediate transitions between 
different nuclear energy levels. In particular, owing to sizeable momentum exchange, dark matter particles can enable de-excitation of nuclear isomers that are extremely long lived with respect to regular radioactive decays. In this paper, we utilize 
data from a past experiment with \nuc{Ta}{180}$^{\rm m}$ 
to search for $\gamma$-lines that would accompany dark matter induced 
de-excitation of this isomer. Non-observation of such transitions above background yields the first direct constraint on the lifetime of \nuc{Ta}{180}$^{\rm m}$ against DM-initiated transitions: $T_{1/2}>1.3\times 10^{14}$~a at 90\% C.I. 
Using this result, we derive novel constraints on dark matter models with strongly interacting relics, and on models with inelastic dark matter particles. Existing constraints are strengthened by this independent new method. The obtained limits are also valid for the Standard Model $\gamma$-decay of \nuc{Ta}{180}$^{\rm m}$.

\end{abstract}

\maketitle

\section{Introduction}
\label{intro}
The existence of massive stable particles with masses commensurate with the electroweak scale is a common feature of many extensions of the Standard Model (SM) \cite{Bertone:2004pz}. Such particles can account for the entirety (or a fraction) of dark matter (DM) in the Universe, motivating intense theoretical and experimental efforts to discover them, or else constrain their properties. Indeed, the searches of weakly interacting massive particles (WIMPs) have progressed to probe tiny cross sections of DM particles 
with nuclei and electrons (see {\em e.g.}\ \cite{Schumann:2019eaa}), and have become the most prominent endeavour in trying to elucidate its nature. 

Nuclear physics, thus far, played a rather limited role in such searches. In most of the large-scale experiments \cite{Aprile:2018dbl,Akerib:2016vxi,Cui:2017nnn,PhysRevD.100.022004}, the nuclear physics input is often limited to refinement of the nuclear matrix elements ({\em e.g.}\ providing a better treatment of elastic nuclear form-factors). Occasionally, ideas with {\em excitation} of nuclear levels by WIMPs has been explored as a way of complementing main searches \cite{Engel:1999kv,Avignone:2000ui,Baudis:2013bba}. 

In recent work \cite{Pospelov:2019vuf}, it was argued that the existence of extremely long lived nuclear isomers can be used as a tool for DM searches, offering a unique probe of DM candidates that are otherwise undetectable in conventional underground experiments. Nuclear isomers are often extremely long-lived, because their conventional $\gamma$-decay would typically require a very significant change of the angular momentum by $L$ units, leading to a strong power suppression of the decay rate by a factor of $\propto (R_Nk_\gamma)^{2L}$ (here $R_N$ is the nuclear radius, and $k_\gamma$ is the wave number of the emitted \gray,
with typical values of $10^{-2}$-to-$10^{-3}$ for $R_Nk_\gamma$). 
It is important to recognize that interaction with a DM particle,
may lead to a direct {\em DM-induced de-excitation} of an isomer ${\cal N}(*) $
to a lower level ${\cal N}(0)$ 
\begin{equation}
\label{de-ex}
{\cal N}(*) + {\rm DM} \to {\cal N}(0) + {\rm DM}, 
\end{equation}
where $\Delta E$, the excess of nuclear energy, is released as 
the kinetic energy of the final state particles. Crucially, this process 
need not be kinematically suppressed, as $(R_Nk_\gamma)^{2L}$ is 
changed to $(R_N\sqrt{2\Delta E \mu}/\hbar)^{2L}$, 
where $\mu$ is a reduced mass
of the DM-nucleus system. The minimum momentum transfer $q_0 = \sqrt{2\Delta E \mu}$ can be as large as $\hbar R_N^{-1}$, so that the main suppression factor can be fully lifted. In this paper, we implement the first direct search of process (\ref{de-ex}), taking the isomeric state to be the famous \nuc{Ta}{180}$^{\rm m}$ nucleus. 

\begin{figure}
\includegraphics[width=0.48\textwidth]{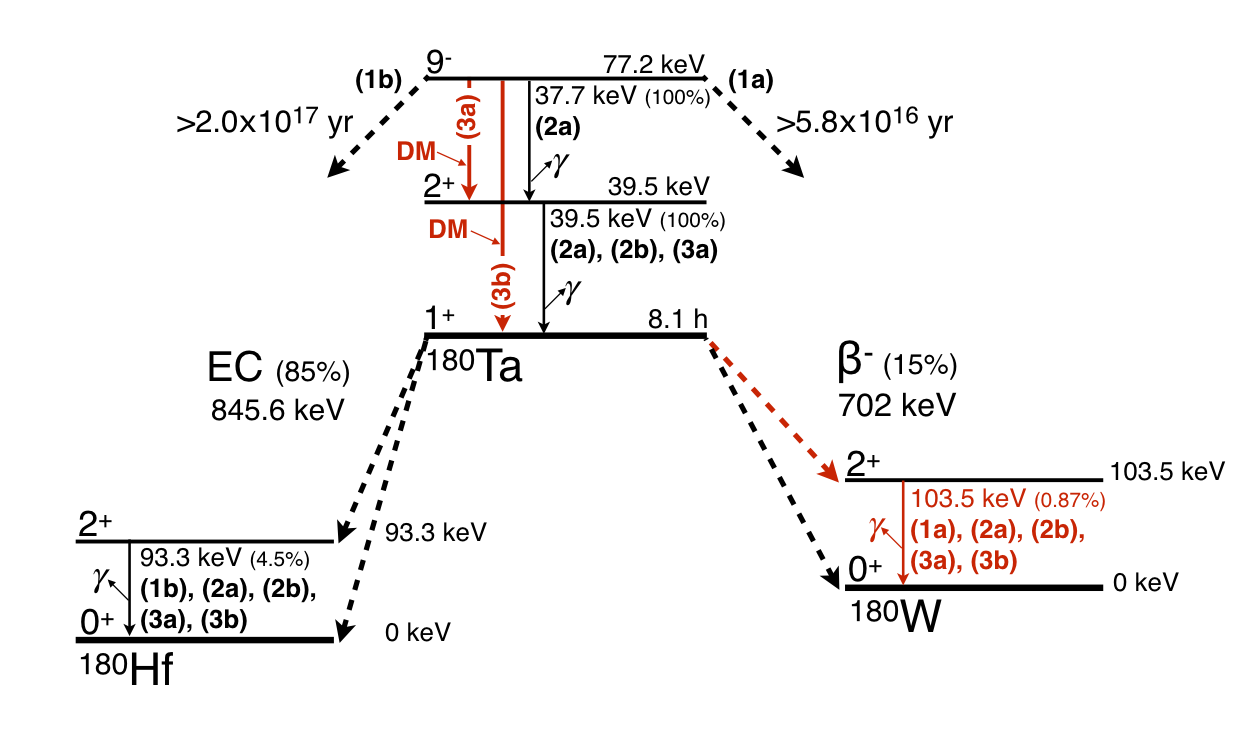}
\caption{\label{fig:decayScheme} Decay scheme of \nuc{Ta}{180}$^{\rm m}$ with data from \cite{NuclData} illustrating the different decay modes: (1a,2a) $\beta^-$ and EC, (2a,2b) $\gamma$-decay and IC, (3a,3b) DM induced decay to $2^+$ and $1^+$ state. The investigated decay modes and signal \gray\ are highlighted in red.}
\label{fig:decayscheme}
\end{figure}

The conventional decay of \nuc{Ta}{180}$^{\rm m}$ was investigated numerous times in the past but has never been observed. The ${\rm E}_7$ transition required for the decay of the $9^-$ isomeric state makes it especially long-lived and remarkably stable on cosmological timescales. Additional interest to this nucleus stems from its relatively large abundance ($\simeq 10^{-4}$ of natural tantalum), which is difficult to reconcile with nucleosynthesis models, where odd-odd nuclei such as \nuc{Ta}{180}$^{\rm m}$  are 
difficult to preserve. 
\fig \ref{fig:decayScheme} shows the possible decay modes of \nuc{Ta}{180}$^{\rm m}$ which are listed below:
The $\beta^-$-decay (1a) and electron capture (EC) (1b) are 3$^{\rm rd}$ forbidden non-unique transitions which were recently investigated in \cite{Lehnert:2016iku}. The experimental signatures are the de-excitation $\gamma$-cascades from the $6^+$ excited states of \nuc{W}{180} and \nuc{Hf}{180}, respectively (not shown). Current half-life limits are at \baseT{5.8}{16}~a for (1a) and \baseT{2.0}{17}~a for (1b) (90\% C.I.) \cite{Lehnert:2016iku} whereas theoretical calculations predict \baseT{5.4}{23}~a for (1a) and \baseT{1.4}{20}~a for (1b) \cite{Ejiri:2017dro,Ejiri:2019ezh}. 
The $\gamma$-decay (2a) and internal conversion (2b) to the \nuc{Ta}{180} ground state is followed by the 8.1~h delayed $\beta^-$ decay or EC into the ground and first excited states of \nuc{W}{180} and \nuc{Hf}{180}. The experimental signatures are the 103.5~keV and 93.3~keV de-excitation \grays\ for (2b) and the additional 37.7~keV and 39.5~keV \grays\ for (2a). Theoretical half-life estimates are \baseT{1.4}{31}~a for (2a) and \baseT{8}{18}~a for (2b). The internal conversion  mode is predicted to be the dominant decay channel in \cite{Ejiri:2017dro,Ejiri:2019ezh}. 

The DM induced \nuc{Ta}{180}$^{\rm m}$ de-excitations $9^- -2^+$ (3a) and $9^- -1^+$ (3b) are the focus of this work. The experimental signature for (3b) is identical to (2b), whereas (3a) has the additional 39.5~keV \gray\ from the \nuc{Ta}{180} $2^+$ de-excitation. 

Currently, the half-life sensitivity for processes (2) and (3) is well below the theoretical prediction of (2b) at \baseT{8}{18}~a, so then a non-observation provides constraints for all decay modes. Should a positive signal eventually be seen, investigation of overburden-dependence will have a discriminating power between natural and DM-induced decays in some models.

Given that only $\sim 10^{-4}$ of natural tantalum can be used for the search, 
it is clear that the search of reaction in \eq \ref{de-ex} cannot compete with massive Xe and Ar-based experiments for constraining tiny elastic cross section. Instead, the \nuc{Ta}{180}$^{\rm m}$ half-life limit will provide a means for probing DM models that cannot be probed with conventional methods \cite{Schumann:2019eaa}. Firstly, if DM interacts strongly with the SM, it will undergo multiple successive collisions 
in the atmosphere and overburden, such that its kinetic energy is reduced far below the threshold for any underground DM detector. But even with vanishingly small incoming velocity, the DM-induced de-excitation of \nuc{Ta}{180}$^{\rm m}$ will go unimpeded. Rocket and balloon experiments with reduced exposures compared to terrestrial direct detection (DD) experiments have been conducted above the atmosphere in order to access DM which has passed through negligible over-burden \cite{Erickcek:2007jv,Hooper:2018bfw}. The results we derive in this 
paper using tantalum allow us to improve on these bounds and strengthen existing constraints with an entirely different method. 
Significant progress is achieved 
in models where strongly-interacting massive particles constitute 
a subdominant component of DM, which is expected in models with thermal DM freeze-out (see {\em e.g.} \cite{DeLuca:2018mzn}).

Another class of WIMP models that could escape conventional searches but be 
discovered/constrained with nuclear isomers are the so-called inelastic DM. In these models, DM is the lighter of two components with a small mass difference, and has dominantly off-diagonal couplings i.e.\ inelastic interactions with the SM. 

Therefore the dominant scattering mechanism requires additional energy for DM excitation, and would not proceed if the mass splitting of the two DM states exceeds available kinetic energy. An addition of nuclear excitation energy $\Delta E$ provided by the isomer in this case, allows accessing larger mass splitting in the DM sector. 
Higgsinos arising from supersymmetric theories are an example of a well motivated inelastic DM candidate that would invoke the WIMP miracle to explain the DM relic abundance, that is not yet ruled out by DM experiments \cite{Bramante:2016rdh}.

\section{Experiment}
\label{Exp}

In this work, we re-analyzed data collected in \cite{Lehnert:2016iku} to look for the signal corresponding to scattering with DM. 
The search in \cite{Lehnert:2016iku} focused on excited state transitions from the EC and $\beta^-$ decays of \nuc{Ta}{180}$^{\rm m}$ which includes higher energy \grays\ up to 332.3~keV. This search focuses on the 93.3~keV and 103.5~keV \grays\ as a signature of the \nuc{Ta}{180} ground state decay. However, the 93.3~keV \gline\ from the EC branch is overlapped with two background \glines\ from \nuc{Th}{234} at $92.38\pm0.01$~keV (2.18\%) and $92.80\pm0.02$~keV (2.15\%), with emission probabilities quoted in parenthesis as well as the Th K-a1 x-ray at 93.31~keV. Thus, only the 103.5~keV \gline\ from the $\beta^-$ branch is used in the analysis as illustrated in red in \fig \ref{fig:decayScheme}. 
A \gline\ at 103.35~keV, also from \nuc{Th}{234}, has 3 orders of magnitude lower emission probability of \baseT{3.2}{-5} and is negligible. A search of nuclear data revealed no reasonable lines which could interfere with this measurement. 
The measurements in \cite{Lehnert:2016iku} were not optimized for such low energy, but nevertheless can be used to set limits. Future experimental improvements are pointed out at the relevant points below and in \sec \ref{sec:exppop}.

The target sample consists of 6 tantalum discs of natural isotopic abundance. The discs have a diameter of 100~mm and a thickness of 2~mm each, summing up to a total mass of 1500.33~g containing 180~mg \nuc{Ta}{180}$^{\rm m}$. 
In total, three measurements are combined: M1 from \cite{Hult:2006exb}, M2 from \cite{Hult:2009cs} and M3 from \cite{Lehnert:2016iku}. Their parameters are condensed in \tab \ref{tab:datasets} for the \gline\ of interest.
All measurements were taken at the HADES underground laboratory with 225~m overburden located at the premise of the Belgium nuclear center SCK$\cdot$CEN in Mol, Belgium. 
Measurement M1 was performed on a single HPGe detector whereas measurement M2 and M3 were performed in a 2-detector sandwich setup. In M2, the data of both detectors were combined into one dataset whereas in M3, they were split into two detectors in order to maximize the advantages of both detector types. 
Ge6 has a 1.0~mm thick copper window and a 0.7~mm thick deadlayer at the top of the Ge-crystal whereas Ge7 has a 1.5~mm thick aluminium-window and a 0.3~$\mu$m deadlayer at the top.
The latter is more suitable for the detection of low energy \grays\ as can be seen in their detection efficiency of the 103.5~keV \gray\ in \tab \ref{tab:datasets}. The slight decrease of detection efficiency from M2 to M3 is due to the growth of the Ge6 dead layer over time.

\begin{table}
\caption{\label{tab:datasets}Overview of the datasets used in the analysis. Columns from left to right denote measurement name, the used HPGe setup, the measurement time, the resolution in FWHM at 103.5~keV and the detection efficiency for the 103.5~keV \gray.}
\begin{ruledtabular}
\begin{tabular}{llrrr}
dataset  & setup & time & FWHM & $\epsilon_{\rm det}$   \\
\hline
M1            & singe det.             & 170~d   & 0.62~keV  & 0.144\% \\
M2            & 2-det sandwich    & 68~d     & 0.70~keV  & 0.239\% \\
M3\_Ge6  & 1-det in sandwich & 176~d  & 0.70~keV  & 0.056\% \\
M3\_Ge7  & 1-det in sandwich & 176~d  & 0.58~keV  & 0.174\% \\
\end{tabular}
\end{ruledtabular}
\end{table} 
\begin{figure*}
\includegraphics[width=0.99\textwidth]{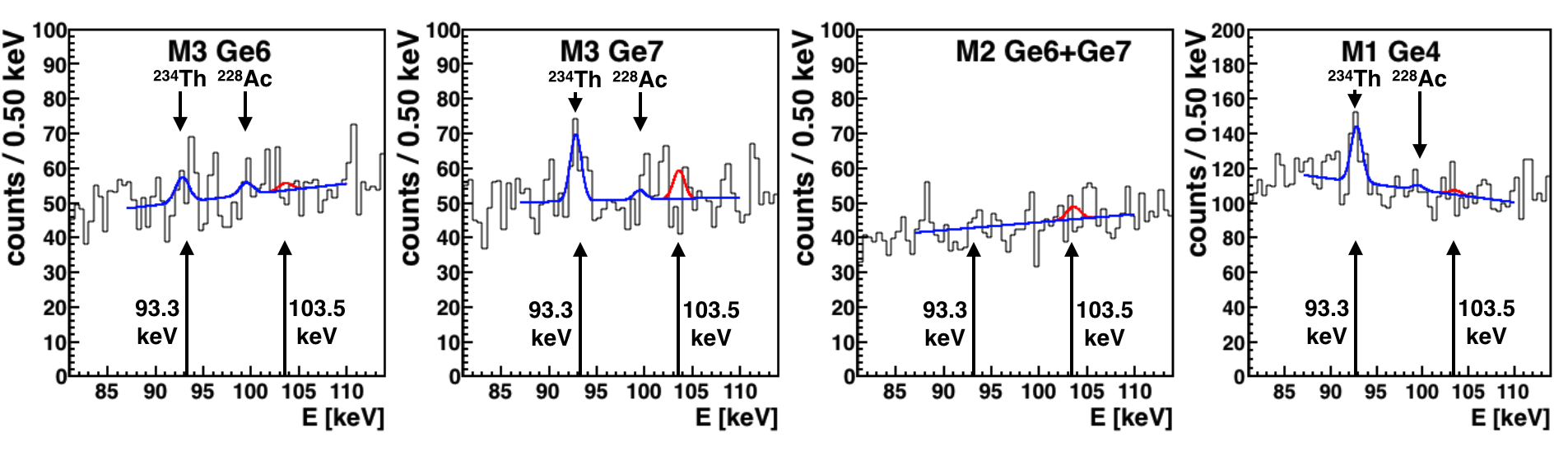}
\caption{\label{fig:pdf_ROI_Composition_Ta180m_bm} Region of interest in each dataset for the 103.5~keV peak search in the $\beta^-$ channel. The best fit is shown in blue and the best fit with the signal peak set to the 90\% C.I.\ half-life limit is shown in red. The arrows indicated the 93.3~keV peak of the EC channel (not used in fit) as well as the named background \glines. }
\end{figure*}

The full energy detection efficiencies were determined with the Monte Carlo code EGSnrc \cite{EGSnrc} 
which has been validated through measurements of reference sources and checked by participation of proficiency tests. The uncertainty is estimated as 10\%. The energy resolutions were routinely determined with a set of calibration sources (\nuc{Am}{241}, \nuc{Cs}{137}, \nuc{Co}{60}) and have an estimated uncertainty of 5\%.

The analysis is a peak search using a Bayesian fit of the Gaussian signal peak on top of a linear background as described in \cite{Lehnert:2016iku}. The fit uses the Bayesian Analysis Toolkit (BAT) \cite{CALDWELL20092197} and includes the inverse half-life as the parameter of interest as well as two parameters to model the background, one parameter for the efficiency, one parameter for the resolution and one parameter for the signal peak position. Three background \glines\ from \nuc{Ac}{228} at 99.51~keV (1.26\%) as well as the two \glines\ from \nuc{Th}{234} mentioned above are included in the fit region of [87,110]~keV and are constrained by two free parameters each, describing their strength and peak position. 
Parameters with known values such as peak positions, the resolution and the detection efficiency have Gaussian priors assigned with the width set to their known uncertainty. This naturally includes systematic uncertainties in the fit. 
The total efficiency is composed of the emission efficiency of the 103.5~keV \gray\ for the \nuc{Ta}{180} ground state decay of $0.87\pm0.24$\% as well as the detection efficiency quoted in \tab \ref{tab:datasets} with 10\% uncertainty. The emission probability dominates to a total uncertainty of 30\% for the efficiency parameter. 
Based on the input parameters, the fit finds zero inverse half-life as the best fit value and hence no signal is observed. The limit setting is based on the marginalized posterior distribution of the inverse half-life of which the 0.9 quantile is used as the 90\% credibility interval (C.I.) at
\begin{equation}
T_{\frac{1}{2}}> 1.3 \times 10^{14}\, {\rm a}\ (90\%\ \rm C.I.)\ .
\label{Thalf}
\end{equation}

Compared to the partial half-life limits for the $\beta$ and EC decay modes of \nuc{Ta}{180}$^{\rm m}$ obtained in \cite{Lehnert:2016iku}, this half-life limit on the \nuc{Ta}{180}$^{\rm m}$ $\gamma$-decay is more than 2 orders of magnitude weaker due to the lower emission probability and detection efficiency of the signal \gray.

\section{Results for Dark Matter}
\label{resdm}
We interpret the partial half-life limits obtained above as limits on DM scattering. The relevant process is DM scattering \nuc{Ta}{180}$^{\rm m}$ ($j=9^-, E=77.2~ {\rm keV}$) to either the lower excited state ($j=2^+,E=39.5~ {\rm keV}$) i.e.\ decay mode (3a), or ground state ($j=1^+,E=0~ {\rm keV}$) i.e.\ decay mode (3b). 

Given $T_{1/2}$, a limit can be set on cross-section for DM $\chi$ with mass $M_\chi$, to scatter off \nuc{Ta}{180}$^{\rm m}$, $\sigma_{\chi {\rm Ta}}$ through,
\begin{equation}
\langle \sigma_{\chi {\rm Ta}} v_\chi \rangle \le \frac{M_{\chi}\log(2)}{T_{\frac{1}{2}} \rho_{\rm lab}} 
\label{halftocross}
\end{equation}
Here $\rho_{\rm lab}$ is the local DM density in the lab. Limits on $\sigma_{\chi {\rm Ta}}$ thus obtained can then be used to set limits on model dependent per-nucleon cross-sections as described next.

\begin{figure*}
\centering

\begin{subfigure}[t]{.45\textwidth}
\centering
\includegraphics[width=\linewidth]{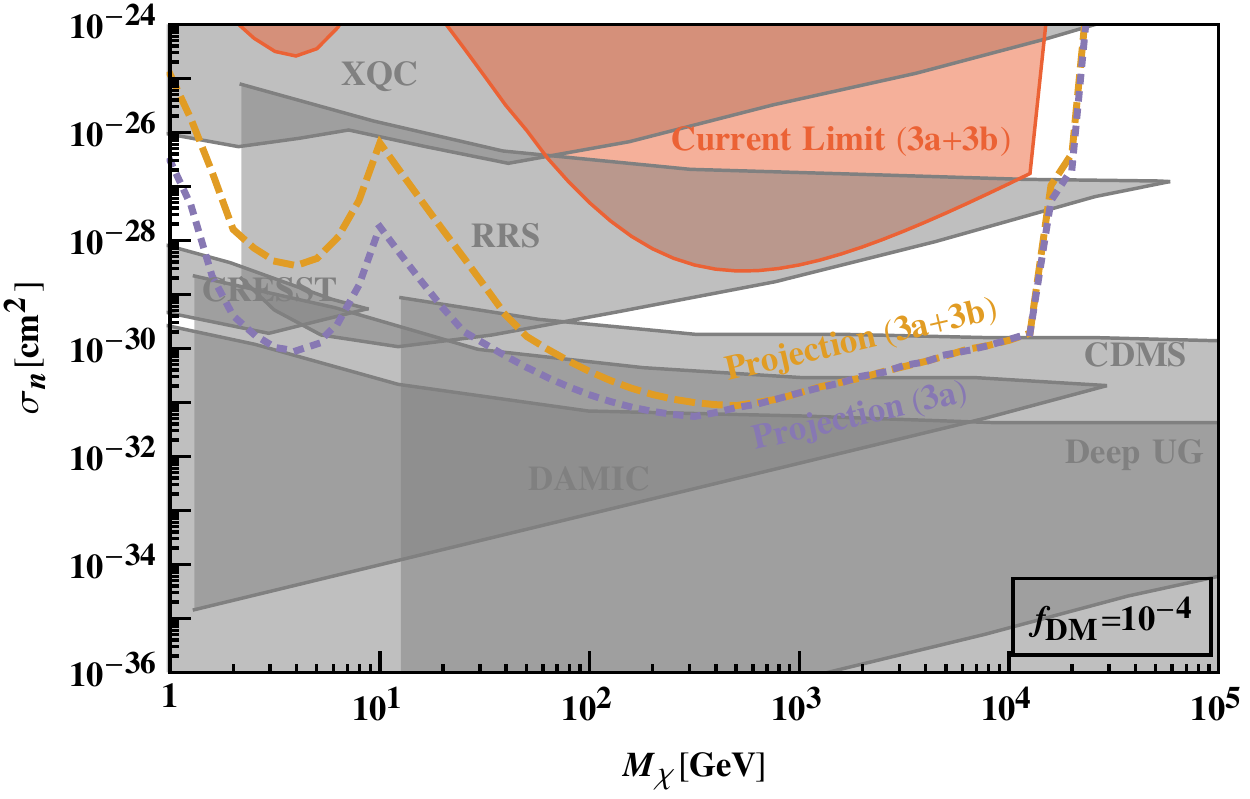}
\end{subfigure}
\begin{subfigure}[t]{.45\textwidth}
\centering

\includegraphics[width=\linewidth]{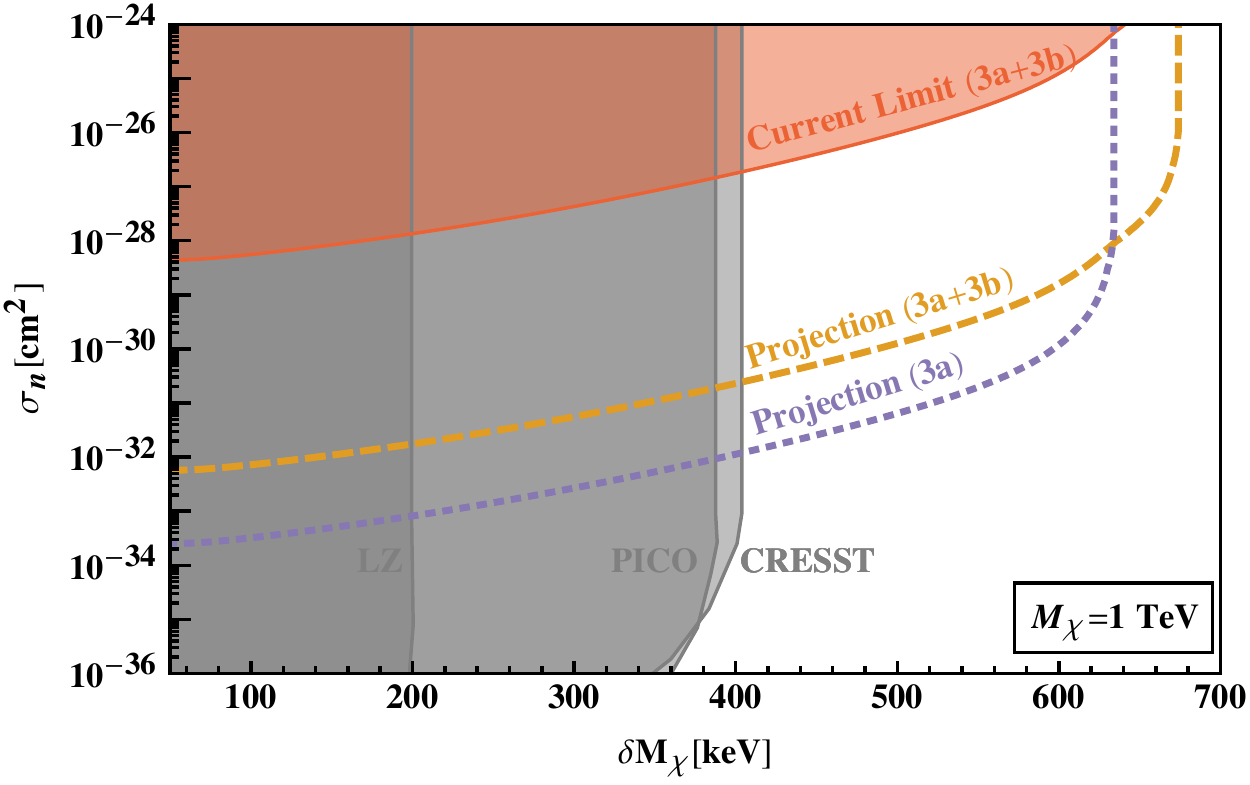}
\end{subfigure}

\caption{\textbf{Left:} 90\% credibility limits on the per-nucleon cross-section for DM that interacts strongly with nuclei from lifetime limit of \nuc{Ta}{180}$^{\rm m}$ corresponding to $T_{1/2} > 1.3 \times 10^{14}$ a are shown in red. $\kappa_L$ is assumed to be 1 (in Eqn. \ref{formfac}). Also shown are limits from existing experiments adapted from \cite{Hooper:2018bfw} in gray. Projections for limits from an experiment that can measure $T_{1/2} > 1 \times 10^{18}$~a in the (3a)+(3b) decay mode are shown in dashed orange and for $T_{1/2} > 4 \times 10^{19}$~a in the (3a) only mode in dashed purple. \textbf{Right: } Limits and projections with the same color coding for inelastic DM with mass splitting $\delta M_{\chi}$. Also shown are limits from existing experiments adapted from \cite{Bramante:2016rdh} in gray.}
\label{smallfigure}
\end{figure*}

\subsection{DM that interacts strongly with nuclei}
DM that interacts strongly with nuclei is poorly constrained due to it its slow down in the atmosphere. The slow down can lead to thermalization with the atmosphere or rock over-burden, and the presence of the gravitational potential eventually leads to a terminal velocity $v_{\rm term}$ towards the center of the earth. The local downward velocity determines the local DM density through flux conservation, $\rho_{\rm lab} = \frac{v_{\rm vir}}{v_{\rm term}} \rho_{\rm ss} $, where $\rho_{\rm lab}$ is the DM density at a location of an underground lab, $\rho_{\rm ss}$ is the solar system DM density, and $v_{\rm vir}$ is the local virial velocity of DM.
The density enhancement $\eta=\frac{\rho_{\rm lab}}{\rho_{\rm ss}}$ was evaluated in \cite{Pospelov:2019vuf} as a function of $\sigma_n$ and $M_\chi$ for the HADES lab with 225 meter overburden. $\eta$ can reach as large as $10^8$ for $M_\chi \sim 100$ GeV and $\sigma_n > 10^{-30}~{\rm cm}^2$. 

For DM lighter than the abundant nuclei on earth, there is an additional trapped thermalized population which was estimated in \cite{Neufeld:2018slx}. We use this in addition to the density estimated in \cite{Pospelov:2019vuf} for limit setting purposes. We provide $\eta$ as a function of $\sigma_n$ and relevant $M_\chi$ in the Supplemental Materials.

We model the cross-section as a generic strong-scale interaction i.e. $\sim \frac{1}{\Lambda_{\rm QCD}^2}$ through the exchange of meson-like hadron resonances, and its reference per-nucleon cross-section is taken to be $\sigma_n$. 
Following \cite{Pospelov:2019vuf} the total cross-section for $\chi$ to scatter off \nuc{Ta}{180}$^{\rm m}$ can be estimated by the following ansatz,

\begin{align}
\langle \sigma_{\chi {\rm Ta}} v_\chi \rangle = \text{Min}\left(\sigma_n \frac{\mu_{{\rm Ta},\chi}}{q_0} ,4\pi R_{\rm Ta}^2\right)\mathcal{S}_f(\bm{q_0}).
\label{finaleqstrong}
\end{align}
Throughout this paper, we use natural units, $\hbar = c =1$. Here, $\mu_{{\rm Ta},\chi}$ is the tantalum-DM reduced mass, $q_0= \sqrt{\Delta E \times \mu_{{\rm Ta},\chi}}$ is the momentum exchange, and $R_{\rm Ta}$ is the radius of tantalum nuclei. The quantity $\mathcal{S}_f(\bm{q_0})$ is the square of the nuclear form-factor that captures the inelastic matrix element for the down-scatter of the isomeric state to one of the lower states. Following \cite{Pospelov:2019vuf} it is estimated from the Weisskopf estimates and includes the hindrance factor $\epsilon_H$ prescribed in \cite{Ejiri:2017dro},
\begin{equation}
 \mathcal{S}_f(\bm{q_0})= \sum_L \kappa_L j_L^2(qR) \epsilon_H
 \label{formfac}
 \end{equation}   
 Here $j_L$ are the spherical Bessel functions. The sum runs over odd L, $7 \le L \le 11$ for $\left(3a\right)$ and $L=9$ for $\left(3b\right)$ scattering processes (Refer Fig.\ref{fig:decayScheme}).  The kludge factor $\kappa_L$ is present to account for deviations from the ansatz that is not captured by the hindrance factors and can be determined by a scattering experiment or observation of the SM decay.

Since it is an exothermic reaction, the counting rate depends on the local DM density and not the flux. 
We can use \eq \ref{finaleqstrong} along with $\eta$ calculated in \cite{Pospelov:2019vuf}, the relation in \eq \ref{halftocross} and the limit in \eq \ref{Thalf} to set limits on $\sigma_n$. This limit will depend on $f_{\rm DM}$, the fraction of solar system DM in $\chi$ particles. Limit contours for $\kappa_L=1$ are plotted in the $\sigma_n$ vs $M_{\chi}$ plane in \fig \ref{smallfigure} (left) for $f_{\rm DM}=10^{-4}$. We also show limits from existing experiments which are adapted from \cite{Hooper:2018bfw}. Stringent new limits are set for $M_\chi>50$ GeV in the strongly interacting regime. There is a drastic reduction in parameter space ruled out by existing DD experiments for such small dark matter fractions. However, \nuc{Ta}{180}$^{\rm m}$ has a much slower drop-off in sensitivity owing to its unique ability to look for slowed down DM that has a large local number density. Limits for different $f_{\rm DM}$ are discussed in the Supplementary Material. 


\subsection{Inelastic Dark Matter}
DM $\chi$ could have dominantly non-diagonal couplings with the SM, i.e.\ scattering relevant to DD is determined by the operator $\mathcal{L}\supset (G_F^\chi)^2  \bar{\chi} \chi' \bar{N}{N}$, where $N$ is the target nucleus, $\chi'$ is non-degenerate with $\chi$ and $\delta M_{\chi} = M_{\chi'}-M_{\chi}$. As a result this energy difference $\delta M_{\chi}$ has to be supplied by the initial kinetic energy in the center of mass frame in conventional DD experiments. Limits from these are summarized in \cite{Bramante:2016rdh} and plotted in \fig \ref{smallfigure} (right), in gray for $M_\chi =1$ TeV. The maximum velocity of the truncated DM velocity distribution, as well as the energy thresholds required for detection, ultimately set the limit on the largest splitting that can be probed, $\delta M_\chi \lesssim 400$ keV in CRESST \cite{cresst} and PICO \cite{amole2016dark}.

The extra energy available for scattering with $^{180\text{m}}$Ta, can uniquely set limits on models with higher energy splittings. The kinematics for this process as well as the estimation of the relevant rates were described in detail in \cite{Pospelov:2019vuf}. The largest splitting that can be accessed with an isomeric transition with energy $\Delta E_N$ is given by, $\delta M_\chi \lesssim \frac{1}{2} \mu_{\chi m_N} (v_\text{E}+v_\text{esc})^2 +\Delta E_N$. Here $v_\text{E}=240$~km/s is the earth velocity in the Milky Way frame and $v_\text{esc}=600$~km/s is the escape velocity, the cut-off of the Maxwell Boltzmann velocity distribution of DM. Unlike in the case of elastic collisions, the overburden does not have stopping power below the center of mass kinetic energy threshold.

Like in the case of strongly interacting DM, $\sigma_{\chi {\rm Ta}}$ can be expressed in terms of the per-nucleon cross-section $\sigma_n$ and relevant kinematic integrals \cite{Pospelov:2019vuf}. We do not repeat this discussion here for brevity. Limits on inelastic DM with mass $M_\chi=1$ TeV for $\kappa_L=1$, as interpreted from the limits on the non-observation of the isomeric transition are displayed in Fig.~\ref{smallfigure} (right) as the shaded red region. This can be compared with existing limits. The large improvement in threshold arises due to the extra energy available in the isomer as well as not requiring a specific range in recoil energy. However, is important to note that inelastic DM models with such large cross-sections are hard to model build \cite{Pospelov:2019vuf}.

\section{Conclusion and Future Experiments}
\label{sec:exppop}

This analysis demonstrates that current nuclear isomer samples can be used to probe strongly interacting DM. Such particles are too slow to be detectable by conventional experiments since they require an ``exothermic'' process to make them experimentally observable. Inelastic DM is another class of models that can be probed by such ``exothermic'' isomers. New parameter space is excluded and existing constraints are strengthened with a new independent method.

The experimental setup can be further optimized for these searches. 
The additional \grays\ of 37.7~keV and 39.5~keV from the \nuc{Ta}{180}$^{\rm m}$ $\gamma$ or DM induced decay would complement the experimental signature and could increase the sensitivity since their emission probability is 100\% per decay. However, such low energies require a dedicated $\gamma$-spectroscopy setup which must be optimized by (1) selecting a suitable HPGe detector technology, (2) use a HPGe cryostat with thin entrance window, (3) optimize the source thickness to reduce self-shielding, and (4) use a tantalum sample enriched in \nuc{Ta}{180} to further increase exposure without increasing self-absorption.

An optimal detector choice are commercially available ultra-low background p-type BEGe (Broad Energy Germanium) detectors in which the n+ contact on the front side is removed to eliminate low energy \gray\ attenuation in the germanium dead layer. The thicker deadlayer on the other sides as well as tantalum sample on the front side would effectively shield this setup from external low energy background \grays.  
The wide energy range of BEGe detectors enables a suitable sensitivity to the $\beta^-$ and EC decays of \nuc{Ta}{180}$^{\rm m}$ allowing a simultaneous search for all decay modes. Commercially available carbon fiber entrance windows of the vacuum cryostat would further maximize low energy detection efficiencies.

Tantalum strongly attenuates \grays\ and only the first millimeters of the sample contribute as active source for the low energy signal. A MC study to optimize the source thickness for a future experiment was performed using MaGe \cite{5876017}. A single BEGe detector (70~mm diameter and 30~mm height) with 0.3~$\mu$m dead layer on the top side was placed in a standard vacuum cryostat with 0.6~mm carbon entrance window. A tantalum disc of equal diameter as the detector and variable thickness was placed 1~mm from the carbon window. The detection efficiency of \grays\ generally decreases with increasing source thickness due to increasing self-absorption but the sensitivity equivalent metric of target mass $\times$ detection efficiency plateaus at a certain thickness. This plateau is energy dependent and was found at thicknesses of about 0.2~mm, 0.8~mm, 3.5~mm, and 7~mm, for 39.5, 103.5, 234.0, and 332.3~keV \grays, respectively. Thus, for the low-energy search of strongly interacting DM, the optimal source thickness is around 1~mm with a total tantalum mass of 64~g. 
This optimized setup would improve the detection efficiency for the 103.5~keV \gray\ from 0.06\% (Ge6) to 0.17\% (Ge7) in this work to about 8.7\% using 24 times less tantalum. A half-life sensitivity of \baseT{3}{14}~a can be achieved for the same peak search within 6 months of measurement using a single detector.
Triggering on the 39.5~keV \gray, the sensitivity increases to \baseT{9}{15}~a for the (2a) and (3a) decay modes due to the higher emission probability. 
Tantalum enriched to 5.5\% \nuc{Ta}{180}$^{\rm m}$ (as used in \cite{PhysRevC.31.1494}) would increase the sensitivity by a factor of about 500 w.r.t.\ to the natural isotopic abundance of 0.012\%.
A simple scaling of the search by running e.g.\ 14 detectors and targets in sandwich configuration (similar to the setup in \cite{Kim2019}) as well as increasing the measurement time to 3 years would increase the sensitivity by another order of magnitude to \baseT{1}{18}~a for the 103.5~keV search (3a)+(3b) and \baseT{4}{19}~a for the 39.5~keV search (3a), respectively. This would also allow to test the theoretical half-life prediction of the $\gamma$-decay mode (2b) from \cite{Ejiri:2017dro}. 
The projected sensitivities for the strongly interacting DM and inelastic DM are shown in \fig \ref{smallfigure} for the (3a)+(3b) type search in dashed orange and for the (3a) type search in dashed purple. The orange lines have larger thresholds due to the additional energy available for the (3b) scatter but lower sensitivity due to the smaller branching ratio of the 103.5~keV \gray.

Different detector technologies could be used to advance this search even further. Large area segmented semi-conductor detectors with thin dead layers could be used to maximize the detection efficiency and background rejection of the low energy \grays\ from a tantalum foil sandwiched in-between them \cite{amman2018,Protic2005}. Another idea is to operate a tantalum crystal as a cryogenic bolometer below 100~mK \cite{Irimatsugawa:2015bi}. In this target=detector approach, the low energy \grays\ do not have to escape the detector and various crystal sizes are possible to fully optimize the signal to background ratio. 

 The search for DM using these isomers could be improved with additional experimental work that can reduce theoretical uncertainties. Theoretical estimates of hindrance factors from \cite{Ejiri:2017dro} were used in this work to account for the hitherto unmeasured transition matrix element. This is an order of magnitude estimation which could be more accurately determined by observing the decay or through scattering with SM projectiles. Until now, photons \cite{photondepop} and neutrons \cite{neutrondepop} have been used to scatter with \nuc{Ta}{180}$^{\rm m}$ to produce de-excitations, albeit the interaction has always gone through a compound nucleus/excited state, since both the photon/neutron can be absorbed. Inelastic scattering with electrons which have been used to estimate transition charge densities for large $\Delta J$ transitions \cite{sandor1993shape}, could be repeated for $^{180{\rm m}}$Ta.
Additionally, the emission probability of the 103.5~keV \gray\ has a large uncertainty of 28\%. While this uncertainty is correctly taken into account in the prior probability of the Bayesian analysis, it would help future experiments to determine the emission probability more precisely.

 Furthermore, as pointed out in \cite{Pospelov:2019vuf}, interesting inelastic DM parameter space can be probed using existing samples of $^{\rm 178m}\text{Hf}$ and $^{\rm 137m}\text{Ba}$ produced as fission waste. These searches require more sophisticated experimental setups, but given the generic nature of the proposed search and the demonstrated feasibility of the approach, we believe that it would be opportune to perform such searches in these isomers.

\section{Acknowledgements}
We would like to thank Alexey Drobizhev and Vivek Singh for discussions on alternative detector technologies. S.R. was supported in part by the NSF under grants PHY-1638509, the Simons Foundation Award 378243 and the Heising-Simons Foundation grants 2015-038 and 2018-0765. H.R. is supported in part by the DOE under contract DE-AC02-05CH11231. 
The work in HADES by Gerd Marissens, Heiko Stroh and Euridice-staff is gratefully acknowledged. The measurements were enabled though the JRC open access initiative to HADES, Project number 21-14.

\bibliographystyle{unsrt}
\bibliography{bibliography}

\clearpage

\onecolumngrid
\appendix
\section*{Supplementary Material}

\label{supp}

\subsection{Overview of decay modes}

\begin{table}[h]
\caption{\label{tab:decayModes} Overview of possible \nuc{Ta}{180}$^{\rm m}$ ($9^-$) decay modes including $\beta^-$-decay, electron capture (EC), $\gamma$-decay, internal conversion (IC), and the newly proposed dark matter induced decay (DM). }
\begin{ruledtabular}
\begin{tabular}{lll}
   \nuc{Ta}{180}$^{\rm m}$ decay modes &  signature \grays\ [keV]  \\
\hline
(1a)  $\beta^-$ \nuc{W}{180} ($6^+$) & 350.9, 234.9, 103.5      \\
(1b)  EC \nuc{Hf}{180} ($6^+$)        & 332.3, 215.3, 93.3       \\
(2a)  $\gamma$ \nuc{Ta}{180} ($2^+$)  & 37.7, 39.5, 93.3, 103.5  \\
(2b)  IC \nuc{Ta}{180} ($2^+$)        & 39.5, 93.3, 103.5  \\
(3a)  DM \nuc{Ta}{180} ($2^+$)        & 39.5, 93.3, 103.5  \\  
(3b)  DM \nuc{Ta}{180} ($1^+$)        & 93.3, 103.5        \\ 
\end{tabular}
\end{ruledtabular}
\end{table}

\subsection{Limits on strongly interacting dark matter fraction}

\begin{figure*}[htpb]
\centering

\begin{subfigure}[t]{.45\textwidth}
\centering
\includegraphics[width=\linewidth]{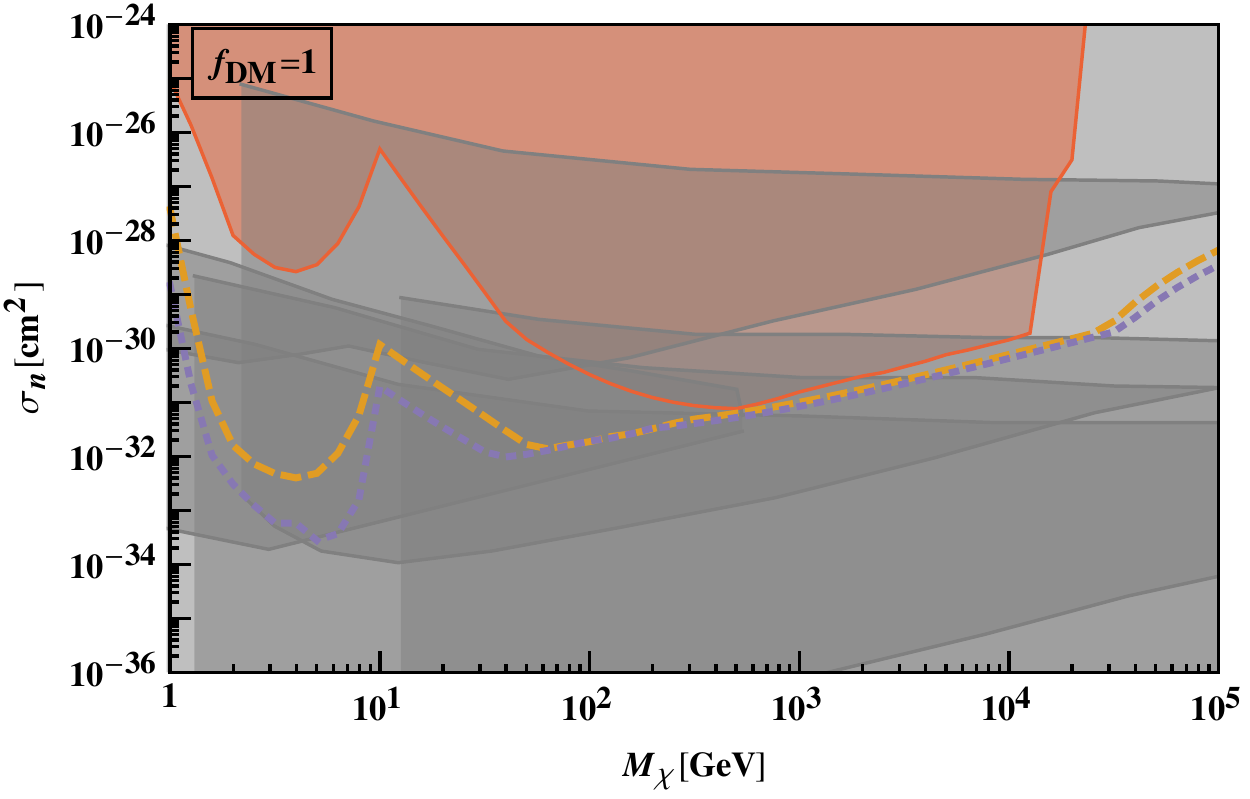}
\end{subfigure}
\begin{subfigure}[t]{.45\textwidth}
\centering

\includegraphics[width=\linewidth]{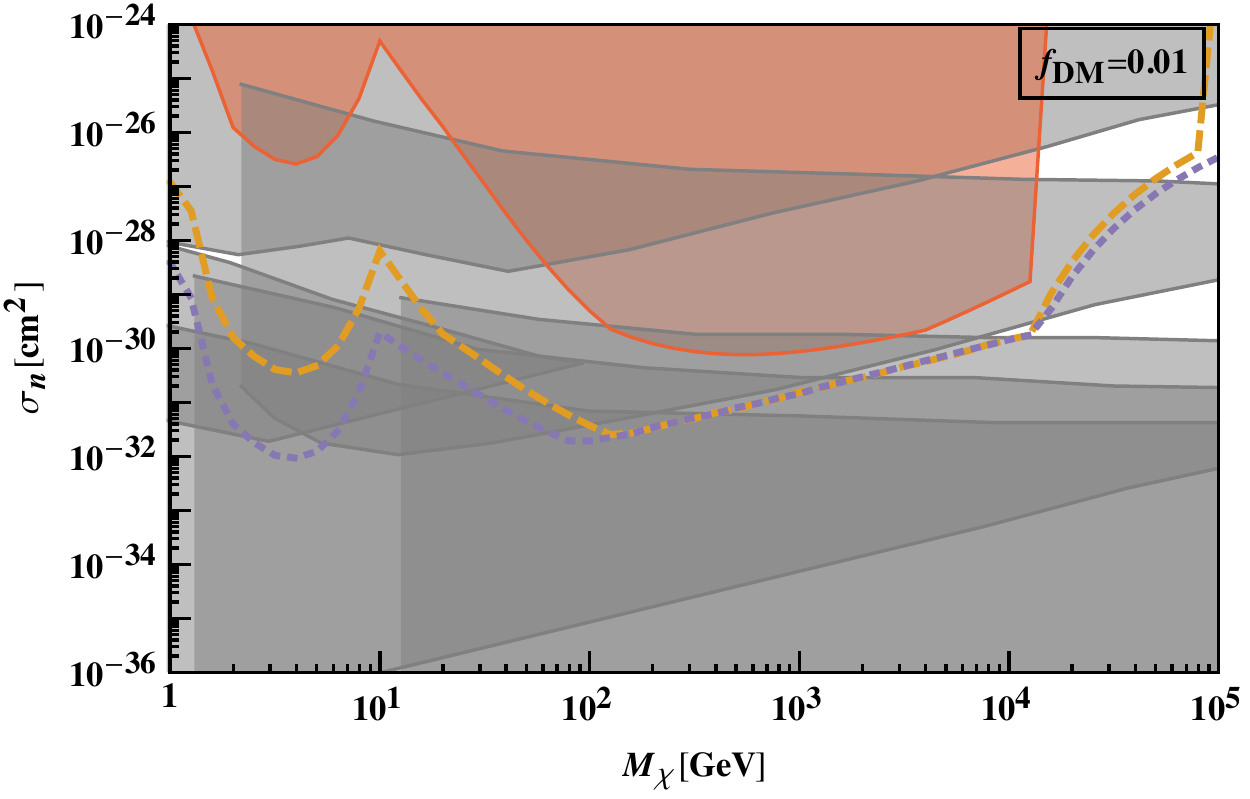}
\end{subfigure}

\medskip

\begin{subfigure}[t]{.45\textwidth}
\centering
\vspace{0pt}
\includegraphics[width=\linewidth]{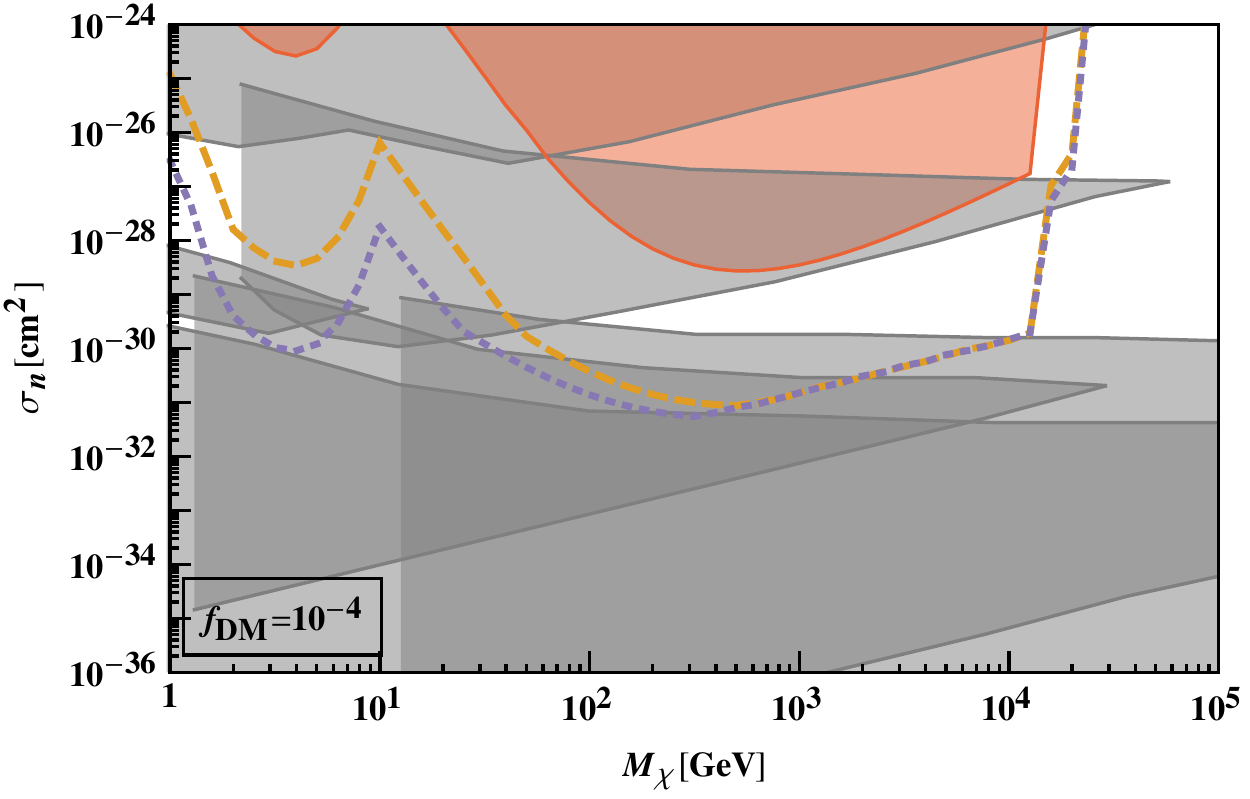}
\end{subfigure}
\begin{subfigure}[t]{.45\textwidth}
\centering
\vspace{0pt}
\includegraphics[width=\linewidth]{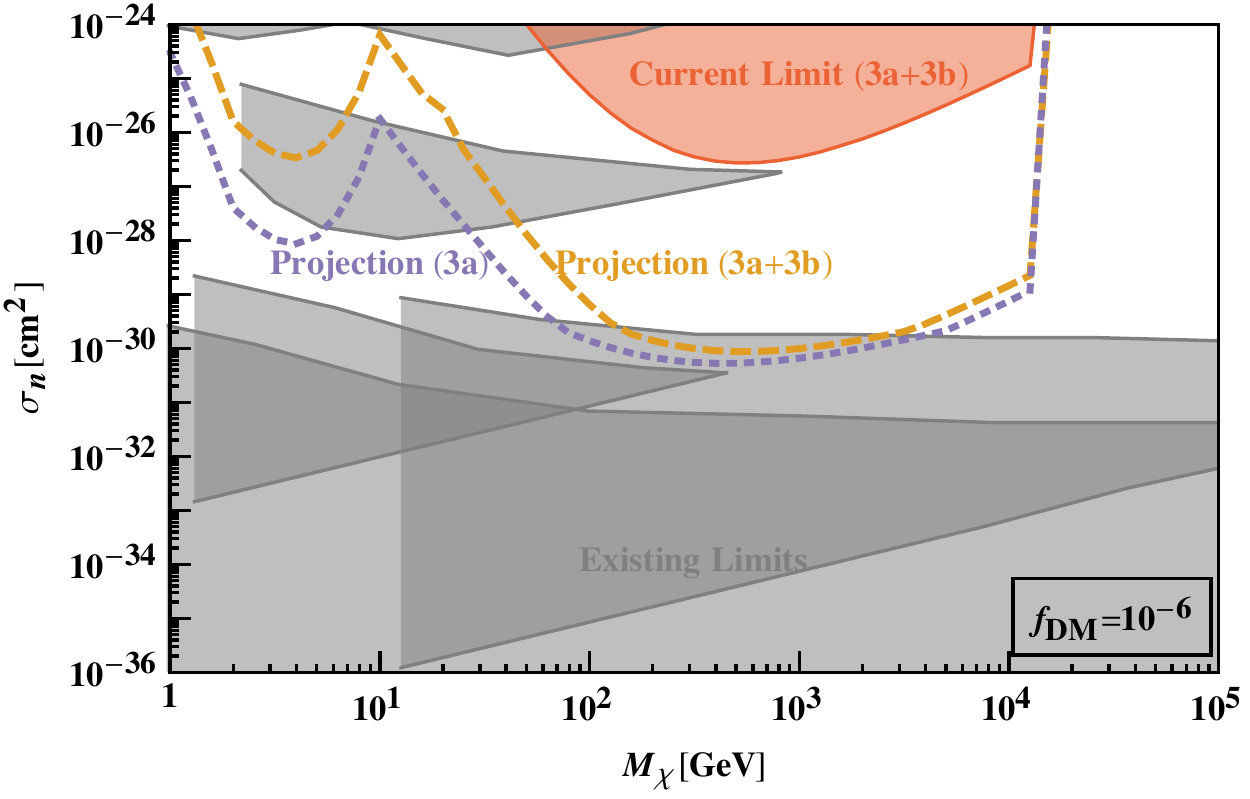}
\end{subfigure}
\medskip
\caption{90\% credibility limits on the per-nucleon cross-section for DM that interacts strongly with nuclei from lifetime limit of \nuc{Ta}{180}$^{\rm m}$ corresponding to $T_{1/2} > 1.3 \times 10^{14}$ a for $\kappa_L=1$ are shown in red. Also shown are limits from existing experiments adapted from \cite{Hooper:2018bfw} in gray. Projections for limits from an experiment that can measure $T_{1/2} > 1 \times 10^{18}$~a in the (3a)+(3b) decay mode are shown in dashed orange and for $T_{1/2} > 4 \times 10^{19}$~a in the (3a) only mode  in dashed purple}
\label{largefigure1}
\end{figure*}

While $f_{\rm DM}=1$ is already adequately exhausted by existing experiments in Fig.~\ref{largefigure1}, especially the satellite based XQC \cite{xqc} and the balloon-based RRS \cite{rrs}, unexplored parameter space opens up even for $f_{\rm DM}=0.01$. Some of this space (around $10$ TeV) is now ruled out through this work.  Stringent limits are also set for $f_{\rm DM}=10^{-4}~{\rm and}~10^{-6}$. As $f_{\rm DM}$ gets more dilute, there is a drastic reduction in parameter space ruled out by existing direct detection (DD) experiments. However, \nuc{Ta}{180}$^{\rm m}$ has a much slower drop-off in cross-section owing to its unique ability to look for slowed down DM that has a large local number density. 

\end{document}

%% file: commands.tex
\newcommand{\nuc}[2]{$^{#2}\rm #1$}

\newcommand{\bb}[1]{$\rm #1\nu \beta \beta$}
\newcommand{\bbm}[1]{$\rm #1\nu \beta^- \beta^-$}
\newcommand{\bbp}[1]{$\rm #1\nu \beta^+ \beta^+$}
\newcommand{\bbe}[1]{$\rm #1\nu \rm ECEC$}
\newcommand{\bbep}[1]{$\rm #1\nu \rm EC \beta^+$}

\newcommand{\rootcern}{\textsc{Root}}
\newcommand{\gerda}{\textsc{Gerda}}
\newcommand{\largeGERDA}{{LArGe}}
\newcommand{\PI}{\mbox{Phase\,I}}
\newcommand{\PIa}{\mbox{Phase\,Ia}}
\newcommand{\PIb}{\mbox{Phase\,Ib}}
\newcommand{\PIc}{\mbox{Phase\,Ic}}
\newcommand{\PII}{\mbox{Phase\,II}}

\newcommand{\geant}{\textsc{Geant4}}
\newcommand{\mage}{\myacs{MaGe}}
\newcommand{\decayzero}{\textsc{Decay0}}

\newcommand{\nPlus}{\mbox{n$^+$ electrode}}
\newcommand{\pPlus}{\mbox{p$^+$ electrode}}

\newcommand{\AOE}{$A/E$}

\newcommand{\order}[1]{\mbox{$\mathcal{O}$(#1)}}

\newcommand{\mul}[1]{\texttt{multiplicity==#1}}

\newcommand{\pic}[5]{
       \begin{figure}[ht]
       \begin{center}
       \includegraphics[width=#2\textwidth, keepaspectratio, #3]{#1}
       \caption{#5}
       \label{#4}
       \end{center}
       \end{figure}
}

\newcommand{\apic}[5]{
       \begin{figure}[H]
       \begin{center}
       \includegraphics[width=#2\textwidth, keepaspectratio, #3]{#1}
       \caption{#5}
       \label{#4}
       \end{center}
       \end{figure}
}

\newcommand{\sapic}[5]{
       \begin{figure}[P]
       \begin{center}
       \includegraphics[width=#2\textwidth, keepaspectratio, #3]{#1}
       \caption{#5}
       \label{#4}
       \end{center}
       \end{figure}
}

\newcommand{\picwrap}[9]{
       \begin{wrapfigure}{#5}{#6}
       \vspace{#7}
       \begin{center}
       \includegraphics[width=#2\textwidth, keepaspectratio, #3]{#1}
       \caption{#9}
       \label{#4}
       \end{center}
       \vspace{#8}
       \end{wrapfigure}
}

\newcommand{\baseT}[2]{\mbox{$#1\cdot10^{#2}$}}
\newcommand{\baseTsolo}[1]{$10^{#1}$}
\newcommand{\THL}{$T_{\nicefrac{1}{2}}$}

\newcommand{\UBI}{$\rm cts/(kg \cdot yr \cdot keV)$}

\newcommand{\Uflux}{$\rm m^{-2} s^{-1}$}
\newcommand{\Ucpd}{$\rm cts/(kg \cdot d)$}
\newcommand{\Uexpo}{$\rm kg \cdot d$}
\newcommand{\UexpoYear}{$\rm kg \cdot yr$}

\newcommand{\UMWE}{m.w.e.}

\newcommand{\Qbb}{$Q_{\beta\beta}$}

\newcommand{\validate}{\textcolor{blue}{\textit{(validate!!!)}}}

\newcommand{\improve}{\textcolor{blue}{\textit{(improve!!!)}}}

\newcommand{\missing}{\textcolor{red}{\textbf{...!!!...} }}

\newcommand{\quanta}{\textcolor{red}{\textit{(quantitativ?) }}}

\newcommand{\misscite}{\textcolor{red}{[citation!!!]}}

\newcommand{\missref}{\textcolor{red}{[reference!!!]}\ }

\newcommand{\PC}{$N_{\rm peak}$}
\newcommand{\BIC}{$N_{\rm BI}$}
\newcommand{\PAPR}{$R_{\rm p/>p}$}

\newcommand{\PCR}{$R_{\rm peak}$}


\newcommand{\gline}{$\gamma$-line}
\newcommand{\glines}{$\gamma$-lines}

\newcommand{\gray}{$\gamma$-ray}
\newcommand{\grays}{$\gamma$-rays}

\newcommand{\bray}{$\beta$-ray}
\newcommand{\brays}{$\beta$-rays}

\newcommand{\aray}{$\alpha$-ray}
\newcommand{\arays}{$\alpha$-rays}

\newcommand{\betas}{$\beta$'s}


\newcommand{\tab}{{Tab.~}}
\newcommand{\eq}{{Eq.~}}
\newcommand{\fig}{{Fig.~}}
\renewcommand{\sec}{{Sec.~}}
\newcommand{\chap}{{Chap.~}}

 \newcommand{\fn}{\iffalse \fi} 
 \newcommand{\tx}{\iffalse \fi} 
 \newcommand{\txe}{\iffalse \fi} 
 \newcommand{\sr}{\iffalse \fi} 

%% file: Tantalum.bbl
\begin{thebibliography}{10}

\bibitem{Bertone:2004pz}
Gianfranco Bertone, Dan Hooper, and Joseph Silk.
\newblock {Particle dark matter: Evidence, candidates and constraints}.
\newblock {\em Phys. Rept.}, 405:279--390, 2005.

\bibitem{Schumann:2019eaa}
Marc Schumann.
\newblock {Direct Detection of WIMP Dark Matter: Concepts and Status}.
\newblock {\em J. Phys.}, G46(10):103003, 2019.

\bibitem{Aprile:2018dbl}
E.~Aprile et~al.
\newblock {Dark Matter Search Results from a One Ton-Year Exposure of XENON1T}.
\newblock {\em Phys. Rev. Lett.}, 121(11):111302, 2018.

\bibitem{Akerib:2016vxi}
D.~S. Akerib et~al.
\newblock {Results from a search for dark matter in the complete LUX exposure}.
\newblock {\em Phys. Rev. Lett.}, 118(2):021303, 2017.

\bibitem{Cui:2017nnn}
Xiangyi Cui et~al.
\newblock {Dark Matter Results From 54-Ton-Day Exposure of PandaX-II
  Experiment}.
\newblock {\em Phys. Rev. Lett.}, 119(18):181302, 2017.

\bibitem{PhysRevD.100.022004}
R.~Ajaj et~al.
\newblock Search for dark matter with a 231-day exposure of liquid argon using
  {DEAP-3600} at {SNOLAB}.
\newblock {\em Phys. Rev. D}, 100:022004, Jul 2019.

\bibitem{Engel:1999kv}
J.~Engel and P.~Vogel.
\newblock {Neutralino inelastic scattering with subsequent detection of nuclear
  gamma-rays}.
\newblock {\em Phys. Rev.}, D61:063503, 2000.

\bibitem{Avignone:2000ui}
F.~T. Avignone, R.~L. Brodzinski, H.~S. Miley, J.~H. Reeves, A.~A. Klimenko,
  S.~B. Osetrov, A.~A. Smolnikov, and S.~I. Vasilev.
\newblock {Results of the pilot experiment to search for inelastic interactions
  of WIMPs with Ge-73}.
\newblock {\em Phys. Atom. Nucl.}, 63:1264--1267, 2000.
\newblock [Yad. Fiz.63,1337(2000)].

\bibitem{Baudis:2013bba}
L.~Baudis, G.~Kessler, P.~Klos, R.~F. Lang, J.~Menéndez, S.~Reichard, and
  A.~Schwenk.
\newblock {Signatures of Dark Matter Scattering Inelastically Off Nuclei}.
\newblock {\em Phys. Rev.}, D88(11):115014, 2013.

\bibitem{Pospelov:2019vuf}
Maxim Pospelov, Surjeet Rajendran, and Harikrishnan Ramani.
\newblock {Metastable Nuclear Isomers as Dark Matter Accelerators}.
\newblock 2019.

\bibitem{Lehnert:2016iku}
B.~Lehnert, M.~Hult, G.~Lutter, and K.~Zuber.
\newblock {Search for the decay of nature's rarest isotope $^{\rm {180m}}$Ta}.
\newblock {\em Phys. Rev.}, C95(4):044306, 2017.

\bibitem{Ejiri:2017dro}
H.~Ejiri and T.~Shima.
\newblock {K-hindered beta and gamma transition rates in deformed nuclei and
  the half-life of $^{\rm {180m}}$Ta}.
\newblock {\em J. Phys.}, G44(6):065101, 2017.

\bibitem{Ejiri:2019ezh}
H.~Ejiri, J.~Suhonen, and K.~Zuber.
\newblock {Neutrino–nuclear responses for astro-neutrinos, single beta decays
  and double beta decays}.
\newblock {\em Phys. Rept.}, 797:1--102, 2019.

\bibitem{NuclData}
Brookhaven National~Laboratory National Nuclear Data~Center.
\newblock Nudat (nuclear structure and decay data), March 18, 2008 2008.

\bibitem{Erickcek:2007jv}
Adrienne~L. Erickcek, Paul~J. Steinhardt, Dan McCammon, and Patrick~C. McGuire.
\newblock {Constraints on the Interactions between Dark Matter and Baryons from
  the X-ray Quantum Calorimetry Experiment}.
\newblock {\em Phys. Rev.}, D76:042007, 2007.

\bibitem{Hooper:2018bfw}
Dan Hooper and Samuel~D. McDermott.
\newblock {Robust Constraints and Novel Gamma-Ray Signatures of Dark Matter
  That Interacts Strongly With Nucleons}.
\newblock {\em Phys. Rev.}, D97(11):115006, 2018.

\bibitem{DeLuca:2018mzn}
Valerio De~Luca, Andrea Mitridate, Michele Redi, Juri Smirnov, and Alessandro
  Strumia.
\newblock {Colored Dark Matter}.
\newblock {\em Phys. Rev.}, D97(11):115024, 2018.

\bibitem{Bramante:2016rdh}
Joseph Bramante, Patrick~J. Fox, Graham~D. Kribs, and Adam Martin.
\newblock {Inelastic frontier: Discovering dark matter at high recoil energy}.
\newblock {\em Phys. Rev.}, D94(11):115026, 2016.

\bibitem{Hult:2006exb}
Mikael Hult, Jo{\"e}l Gasparro, Gerd Marissens, Patric Lindahl, Uwe W{\"a}tjen,
  Peter~N Johnston, Cyriel Wagemans, and Matthias K{\"o}hler.
\newblock {Underground search for the decay of Ta180m}.
\newblock {\em Physical Review C}, 74(5):054311--5, November 2006.

\bibitem{Hult:2009cs}
Mikael Hult, J.S.~Elisabeth Wieslander, Gerd Marissens, Jo{\"e}l Gasparro, Uwe
  W{\"a}tjen, and Marcin Misiaszek.
\newblock {Search for the radioactivity of 180mTa using an underground HPGe
  sandwich spectrometer}.
\newblock {\em Applied Radiation and Isotopes}, 67(5):918--921, May 2009.

\bibitem{EGSnrc}
I.~Kawrakow, E.~Mainegra-Hing, D.~W.~O. Rogers, F.~Tessier, and B.~R.~B.
  Walters.
\newblock {The EGSnrc code system: Monte Carlo simulation of electron and
  photon transport}.
\newblock {\em National Research Council Canada}, PIRS-701, 2015.

\bibitem{CALDWELL20092197}
Allen Caldwell, Daniel Kollár, and Kevin Kröninger.
\newblock Bat – the bayesian analysis toolkit.
\newblock {\em Computer Physics Communications}, 180(11):2197 -- 2209, 2009.

\bibitem{Neufeld:2018slx}
David~A. Neufeld, Glennys~R. Farrar, and Christopher~F. McKee.
\newblock {Dark Matter that Interacts with Baryons: Density Distribution within
  the Earth and New Constraints on the Interaction Cross-section}.
\newblock {\em Astrophys. J.}, 866(2):111, 2018.

\bibitem{cresst}
G~Angloher, A~Bento, C~Bucci, L~Canonica, X~Defay, A~Erb, F~von Feilitzsch,
  N~Ferreiro Iachellini, P~Gorla, A~G{\"u}tlein, et~al.
\newblock Results on light dark matter particles with a low-threshold cresst-ii
  detector.
\newblock {\em The European Physical Journal C}, 76(1):25, 2016.

\bibitem{amole2016dark}
C~Amole, M~Ardid, DM~Asner, D~Baxter, E~Behnke, P~Bhattacharjee, H~Borsodi,
  Manuel Bou-Cabo, SJ~Brice, D~Broemmelsiek, et~al.
\newblock Dark matter search results from the pico-60 cf 3 i bubble chamber.
\newblock {\em Physical Review D}, 93(5):052014, 2016.

\bibitem{5876017}
Melissa Boswell, Yuen-Dat Chan, Jason~A Detwiler, Padraic Finnerty, Reyco
  Henning, Victor~M Gehman, Rob~A Johnson, David~V Jordan, Kareem Kazkaz,
  Markus Knapp, et~al.
\newblock Mage-a geant4-based monte carlo application framework for
  low-background germanium experiments.
\newblock {\em IEEE Transactions on Nuclear Science}, 58(3):1212--1220, 2011.

\bibitem{PhysRevC.31.1494}
J.~B. Cumming and D.~E. Alburger.
\newblock Search for the decay of $^{180}\mathrm{Ta}^{\mathrm{m}}$.
\newblock {\em Phys. Rev. C}, 31:1494--1498, Apr 1985.

\bibitem{Kim2019}
GW~Kim, SY~Park, IS~Hahn, YD~Kim, MH~Lee, DS~Leonard, EK~Lee, WG~Kang, E~Sala,
  and V~Kazalov.
\newblock Simulation study for the half-life measurement of 180m ta using hpge
  detectors.
\newblock {\em Journal of the Korean Physical Society}, 75(1):32--39, 2019.

\bibitem{amman2018}
Mark Amman.
\newblock Optimization of amorphous germanium electrical contacts and surface
  coatings on high purity germanium radiation detectors, 2018.

\bibitem{Protic2005}
D.~Protic, E.L. Hull, T.~Krings, and K.~Vetter.
\newblock Large-volume {Si(Li)} orthogonal-strip detectors for
  compton-effect-based instruments.
\newblock {\em IEEE Transactions on Nuclear Science}, 52:3181, Dec 2005.

\bibitem{Irimatsugawa:2015bi}
T~Irimatsugawa, S~Hatakeyama, M~Ohno, H~Takahashi, C~Otani, and T~Maekawa.
\newblock {High Energy Gamma-Ray Spectroscopy Using Transition-Edge Sensor With
  a Superconducting Bulk Tantalum Absorber}.
\newblock {\em IEEE Transactions on Applied Superconductivity}, 25(1):1--3,
  June 2015.

\bibitem{photondepop}
SA~Karamian, CB~Collins, JJ~Carroll, and J~Adam.
\newblock Isomeric to ground state ratio in the 180 ta m ($\gamma$, $\gamma$)
  180 ta m, g reaction.
\newblock {\em Physical Review C}, 57(4):1812, 1998.

\bibitem{neutrondepop}
SA~Karamian, CB~Collins, JJ~Carroll, J~Adam, AG~Belov, and VI~Stegailov.
\newblock Fast neutron induced depopulation of the 180 ta m isomer.
\newblock {\em Physical Review C}, 59(2):755, 1999.

\bibitem{sandor1993shape}
RKJ Sandor, HP~Blok, M~Girod, MN~Harakeh, CW~de~Jager, V~Yu Ponomarev, and
  H~de~Vries.
\newblock Shape transition of 146nd deduced from an inelastic
  electron-scattering experiment.
\newblock {\em Nuclear Physics A}, 551(3):378--408, 1993.

\bibitem{xqc}
Dan McCammon, R~Almy, E~e 1~al Apodaca, W~Bergmann Tiest, W~Cui, S~Deiker,
  Massimiliano Galeazzi, M~Juda, A~Lesser, T~Mihara, et~al.
\newblock A high spectral resolution observation of the soft x-ray diffuse
  background with thermal detectors.
\newblock {\em The Astrophysical Journal}, 576(1):188, 2002.

\bibitem{rrs}
J~Rich, R~Rocchia, and Michel Spiro.
\newblock A search for strongly interacting dark matter.
\newblock {\em Physics Letters B}, 194(1):173--176, 1987.

\end{thebibliography}
